%% file: Fernandez-Zapico.Izadi.ea.ITSC26.tex
\documentclass[letterpaper, 10 pt, conference]{ieeeconf} 
\IEEEoverridecommandlockouts
\usepackage[english]{babel}
\usepackage{cite}
\usepackage{amsmath,amssymb,amsfonts}
\usepackage{algorithmic}
\usepackage{graphicx}
\usepackage{subcaption}
\usepackage{textcomp}
\usepackage{amsmath}
\usepackage[dvipsnames]{xcolor}
\usepackage{units}
\usepackage{booktabs}
\usepackage{comment}
\usepackage[hidelinks]{hyperref}
\usepackage[normalem]{ulem}
\usepackage{float}
\usepackage{url}
\usepackage{dsfont}
\usepackage{booktabs,tabularx}
\newtheorem{thm}{Theorem}
\newtheorem{prob}[thm]{Problem} 

\newif\ifmargincomments 
\margincommentstrue
\usepackage{marginnote}

\ifmargincomments

\else

\fi

\title{\LARGE \bf Optimal Control Strategies for a Network of Electric Vehicle Charging Energy Hubs with Smart Scheduling via Distributed Optimization} 

\author{Diego Fernández-Zapico, Finn Vehlhaber, Maedeh Izadi, Theo Hofman, Mauro Salazar
\thanks{Control Systems Technology section, Department of Mechanical Engineering, Eindhoven University of Technology, Eindhoven, The Netherlands. E-mails:  {\tt\small \{d.fernandez.zapico, f.n.vehlhaber, m.izadi.najafabadi1, t.hofman, m.r.u.salazar\}@tue.nl}}%
}

\begin{document}
\maketitle
\begin{abstract}
 This paper studies the cost-optimal operation of a network of charging energy hubs for electric vehicles, which provide onsite renewable energy sources and stationary battery storage and are connected with each other via DC-lines as well as with the distribution grid.
Specifically, we first formulate a dynamic optimal control problem for the entire network as a convex quadratic program, whereby the charging power profiles of the individual vehicles and the energy flows between hubs and the grid are subject to optimization.
Second, we propose a problem decomposition that allows for a distributed solution via ADMM algorithms that preserves global optimality guarantees and privacy of the individual stations.
We showcase our framework on a case-study for the Netherlands considering a two-day ahead deterministic formulation with perfect foresight.
Our results show that compared to the case where charging powers are fixed a priori, optimizing their profiles (V1G) can significantly reduce the operational costs and emissions by more than 25\%.
Moreover, we verify our distributed algorithm against a centralized solution, paving the way to the optimal operation of large networks and online implementations.
\end{abstract}

\vspace{-7pt}
\input{Sections/introduction.tex}
\vspace{-7pt}
\input{Sections/methodology.tex}

\vspace{-7pt}
\input{Sections/results.tex}

\vspace{-6pt}
\input{Sections/conclusions.tex}
\vspace{-4pt}
\section*{Declaration on Generative AI}
\vspace{-4pt}
The authors used LLMs to improve the clarity and structure of some sections in this work.

\bibliographystyle{IEEEtran}
\vspace{-4pt}
 \bibliography{SML_Papers.bib, energyhubs.bib}
 
\end{document}

%% file: Sections/introduction.tex
\section{Introduction}

\begin{figure}[t]
	\centering
	\includegraphics[width=0.7\linewidth]{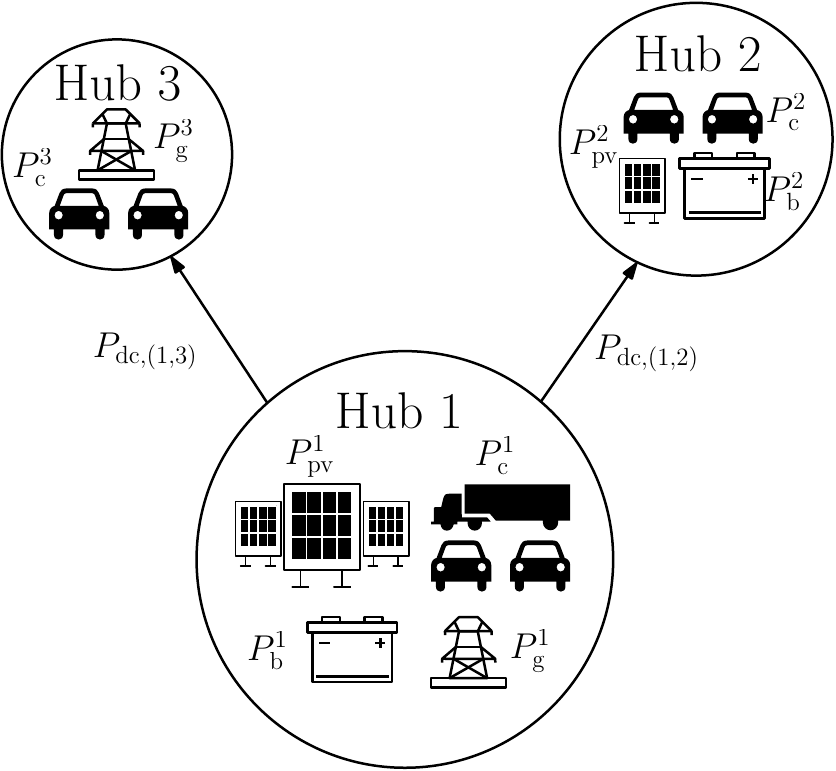}
	\vspace{2pt}
	\caption{Network of three Charging Energy Hubs composed of different components and linked by DC power interconnections providing positive power flows $P_\mathrm{dc}$.}
	\label{fig:network}
	\vspace{10pt}
\end{figure}

Government policy and climate goals support ongoing electrification of transport systems and the development of renewable energy sources. Realizing this transition requires overcoming the challenges of operating and designing such charging infrastructure for road, maritime and air mobility \cite{VehlhaberSalazar2023b, BertucciHofmanEtAl2025}. In the context of road mobility, Charging Energy Hubs (CEHs) are able to increase the flexibility of electric vehicle (EV) charging stations, complementing them with onsite renewable energy sources and battery energy storage systems (BESS) \cite{FernandezZapicoHofmanEtAl2025, IzadiFernandezZapicoEtAl2026}.
Networks of interconnected CEHs, as the one shown in Fig.~\ref{fig:network}, provide additional flexibility to transfer power between the stations, however, they also introduce the challenge of coordinating coupled hubs with privacy concerns on information sharing.
Therefore, both for design and operational purposes there is a need to study this system in an optimization framework.

This paper presents a framework to optimize the operation of a network of CEHs that can be equipped with onsite renewable generation, a BESS and can be connected to the power grid, in terms of power flows among hubs as well as EV charging profiles (also denoted as V1G, which differs from V2G by not allowing EV discharge), and provides a decomposed problem formulation that can be solved with distributed optimization methods.
 
 \textit{Related literature:}
 Centralized control of networks relies on the collection of all component information to compute decisions. In \cite{LeFlochBansalEtAl2019}, authors propose a centralized hierarchical model predictive control strategy for load shaping and voltage regulation in distribution grids equipped with shapeable EV loads, deferrable fixed loads and BESS.
 While authors ensure network feasibility under plug-in and plug-out events, this centralized approach faces privacy and scalability limitations as the network size increases.
 To overcome this, decentralized control can be used as a distributed approach that separates the problem among multiple agents coordinated through limited information sharing~\cite{MotlaghOladigboluEtAl2025}. In this context, the Alternating Direction Method of Multipliers (ADMM) is a distributed method that has become a widely adopted approach in large-scale distributed EV charging optimization for multiple operational goals (cost minimization, voltage regulation) and at multiple hierarchical levels (EV, aggregator, hub) \cite{NimalsiriMediwaththeEtAl2019}.
 The authors of~\cite{RiveraWolfrumEtAl2013} propose a decentralized EV fleet charging framework applying ADMM at the EV level with an aggregator-level coordinator. It enforces consistency between aggregate and individual charging profiles through iterative incentive signals, enabling valley filling and price-based charging.
 However, it performs the decomposition at the EV level within a hierarchical aggregator structure, rather than across interconnected charging stations. 
In~\cite{KianiSheshyekaniEtAl2024} a hierarchical distributed framework is presented for EV charging power optimization coordinated across the grid operator, EV aggregators, and individual EVs, with the goal of EV charging costs minimization, aggregator peak load shaving and grid voltage regulation.  Its main novelty lies in a single-loop ADMM structure that enables all layers to update their decision variables simultaneously within a unified iterative process.
 In \cite{ZhouZouEtAl2021}, a decentralized fast-charging coordination framework is proposed that incorporates battery degradation, user satisfaction, and charging-time minimization via a sparsity-promoting formulation, and is solved using a hierarchical ADMM-based algorithm with convergence guarantees. However, in both \cite{KianiSheshyekaniEtAl2024} and \cite{ZhouZouEtAl2021}, the focus remains on EV-level coordination within distribution networks, without considering integrated energy hubs with local generation, storage, or inter-hub power exchange. Authors in  \cite{MansouriNematbakhshEtAl2024}, propose a robust ADMM-enabled decentralized framework for coordinating community microgrids in local energy and flexibility markets. The architecture consists of three levels: smart residential buildings, where EV charging, flexible loads, and distributed resources are aggregated and modeled as virtual energy storage systems; community microgrids that schedule local generation and storage; and a distribution system operator that clears the market and enforces grid constraints. The framework achieves coordination using adaptive ADMM within a market equilibrium formulation to preserve privacy and ensure network feasibility. In \cite{NasiriZeynaliEtAl2024}, a decentralized peer-to-peer energy trading framework is proposed among networked microgrids, smart parking lots, and the distribution network using an ADMM-based structure with nodal electricity pricing and distributionally robust optimization to handle renewable uncertainty. However, the last two cited papers focus on market-based trading within distribution networks and do not consider physically interconnected CEHs with direct inter-hub power exchange. In \cite{YanChen2023}, authors propose an improved ADMM method for online control of a CEH network, distributed at the hub level and aiming for consensus on energy sharing. However, authors model the EV charging demand as an aggregate which does not allow for shapeable charging profiles, which may be suboptimal.
 
 In conclusion, to the best of our knowledge, no existing study addresses the distributed coordination of physically interconnected CEHs by simultaneously optimizing external power sharing among hubs and  internal power management, including shaping EV power loads.

\textit{Statement of contributions:}
This work proposes a convex optimal control formulation for DC-interconnected CEHs optimizing over EV charging power profiles and the power exchanges among hubs and the main grid. Moreover, we derive an ADMM-based hub-level distributed problem that aims for consensus over the inter-hub power exchanges with limited information sharing, enabling for privacy-aware coordination and large network applications.

\textit{Organization:} The remainder of this paper is structured as follows: Section II describes the  CEH Network model. Section III presents the centralized formulations of the optimization problem. Section  IV explains the formulation of the distributed optimization problem and its solution using ADMM. Section V reports the results of a case study for the Netherlands. Finally, Section VI includes the conclusions and future work.

\paragraph*{Notation} Throughout the paper we use bold letters to denote time series as row vectors, e.g., $\mathbf{v} = \{v[k]\}_{k=1}^N = \left[v_1,\dots,v_N\right]$. A column vector of all ones of appropriate dimension is denoted $\mathds{1}$.

%% file: Sections/methodology.tex
\section{Charging Energy Hub Network}\label{model}
The Charging Energy Hub Network is a group of DC-interconnected hubs as depicted in Fig.~\ref{fig:network}. We model the network as a directed graph with vertices $i \in~\mathcal{H}=~\{1,...,N_\mathrm{h}\}$ corresponding to the hubs, and edges $c=~(i,h) \in~\mathcal{C}$ as their interconnections. Each hub consists of an EV charging station with a power demand $P_\mathrm{c}^i$ sustained by the power from its available components: BESS ($P_\mathrm{b}^i$), solar panels ($P_\mathrm{pv}^i$) and power grid connection ($P_\mathrm{g}^i$). We work in discrete time with index $k \in~\mathcal{K}=~\{0, ..., N_\mathrm{w}-1\}$, with the window size $N_\mathrm{w}$ and time resolution $\Delta t$. At each step $k$ the power balance of the energy hub $i$ is satisfied:
\begin{equation}\label{eq:Pbalance}
    P_{\mathrm{c}}^{i} [k]= P_{\mathrm{g}}^i [k]+ P_{\mathrm{pv}}^i[k] + P_{\mathrm{b}}^i[k] + \sum_{c:i\in c} B_c^i P_{\mathrm{dc},c}[k]
\end{equation}
where $B_c^i$ is the element of the incidence matrix corresponding to the node-edge pairing, and  $P_{\mathrm{dc},c}$ is the power at DC-connection $c$, assuming a lossless connection, limited by
\begin{equation}\label{eq:Pdclims}
	{P}_{\mathrm{dc}, c} [k]\in \left[\underline{P}_{\mathrm{dc}}, \overline{P}_{\mathrm{dc}}\right] \quad \forall k \in \mathcal{K},\;\forall c \in \mathcal{C}.
\end{equation}

\subsection{BESS}
We model the BESS losses using a constant efficiency $\eta$, and the internal battery power as
\begin{equation*}
	P_{\mathrm{ib}}^i [k]= 
	\begin{cases}
		\frac{1}{\eta} \cdot P_{\mathrm{b}}^i [k] & \text{if } P_{\mathrm{b}}^i [k] \ge 0,\\ 
		\eta \cdot P_{\mathrm{b}}^i [k]& \text{otherwise},
	\end{cases} \quad \forall i \in \mathcal{H}, \; \forall k \in \mathcal{K}, 
\end{equation*}
where $P_{\mathrm{b}}^i$ is the external battery power. This constraint can be relaxed to its epigraph, introducing convexity:
\begin{equation}\label{eq:pbloss_relaxed}
	P_{\mathrm{ib}}^i [k]\geq 
	\begin{cases}
		\frac{1}{\eta} \cdot P_{\mathrm{b}}^i[k], \\ 
		\eta \cdot P_{\mathrm{b}}^i[k].
	\end{cases}
\end{equation}
The dynamics of the battery are a discrete linear expression
\begin{equation}\label{eq:EbDyn}
    E_{\mathrm{b}}^i[k\!+\!1] = E_{\mathrm{b}}^i[k] - \Delta t \cdot P_{\mathrm{ib}}^i[k],
\end{equation}
where the initial value is set to the minimum ($\underline{E}_{\mathrm{b}}$)
\begin{equation}\label{eq:EbInit}
	E_{\mathrm{b}}^i[0] = \underline{E}_\mathrm{b}^i.
\end{equation} 
Additionally, the battery is constrained to operate between the minimum and maximum ($\overline{E}_{\mathrm{b}}$) operational levels
\begin{equation}\label{eq:Eblims}
    E_{\mathrm{b}}^i[k] \in \left[\underline{E}_{\mathrm{b}}^i, \overline{E}_{\mathrm{b}}^i\right],
\end{equation}
and, similarly, the output power from the battery is constrained by
\begin{equation}\label{eq:Pblims}
	{P}_{\mathrm{b}}^i[k] \in \left[\underline{P}_{\mathrm{b}}^i, \overline{P}_{\mathrm{b}}^i\right] \quad \forall i\in\mathcal{H},\;\forall k\in\mathcal{K}.
\end{equation}

\subsection{Optimized EV Charging Power}
At a hub, each charging EV constitutes a charging session $j \in \mathcal{S}^i=\{ 1, ..., N_\mathrm{ev}^i\}$ with an energy request $E_{\mathrm{c},j}^i $ that must be completely met within the scheduled session. We optimize the EV charging power $\mathbf{P}_{\mathrm{ev},j}^i\in \mathbb{R}^{1 \times N_\mathrm{w}}$ within the session, for which we assume perfect charging with no losses, such that
\begin{equation}\label{eq:EcOpt}
	E_{\mathrm{c},j}^i = \mathbf{P}_{\mathrm{ev},j}^i \mathds{1} \cdot \Delta t \quad \forall j \in \mathcal{S}^i,
\end{equation}
where the charging power at every time step is constrained by
\begin{equation}\label{eq:PevLimits}
	 {P}_{\mathrm{ev}, j}^i[k] \in \left[\underline{P}_{\mathrm{ev}}, \overline{P}_{\mathrm{ev}}\right] \quad \forall j \in \mathcal{S}^i,\, \forall k \in \mathcal{K},\, \forall i \in \mathcal{H}.
\end{equation}

We enforce charging to take place during the scheduled connection ($k_{0, j}^i$) and disconnection ($k_{\mathrm{f}, j}^i$) time step by constraining
\begin{equation}\label{eq:EVschedule}
    (1-\mathbf{a}_j^i)\odot\mathbf{P}_{\mathrm{ev},j}^i = 0 \quad \forall j \in \mathcal{S}^i, \; \forall i \in \mathcal{H},
\end{equation}
where  $\odot$ is the Hadamard product, $\mathbf{a}_j^i \in \mathbb{R}^{1 \times N_\mathrm{w}}$ is the row in the availability matrix corresponding to the $j$-th session, each element of which can be defined for each step $k$ as:
\begin{equation*}
    a_{j}^i [k] = \begin{cases}
        1  & \text{if } k \in [k_{0, j}^i, k_{\mathrm{f}, j}^i] \quad \forall j  \in \mathcal{S}^i, \\
        0 & \text{otherwise}.
    \end{cases} 
\end{equation*}

We define the total charging power at each hub as
\begin{equation}\label{eq:chargingPower}
	P_{\mathrm{c}}^{i} [k]= \sum_{j\in\mathcal{S}^i} {P}_{\mathrm{ev}, j}^i [k]\quad \forall i \in \mathcal{H}, \; \forall k \in \mathcal{K} .
\end{equation}

\subsection{Power Grid}
We model the power exchanges with the grid to allow buying and selling based on the hourly day-ahead electricity price ($p_{\mathrm{da}}$) under the assumption of perfect efficiency. We define the electricity cost as
\begin{equation*}
    C_{\mathrm{el}}^i [k]= 
    \begin{cases}
    p_{\mathrm{da}}[k]\cdot\Delta t\cdot P_{\mathrm{g}}^i [k]& \text{if } P_{\mathrm{g}}^i[k] \ge 0\\ 
    R_\mathrm{sell} \cdot p_{\mathrm{da}}[k]\cdot \Delta t\ \cdot P_{\mathrm{g}}^i[k] & \text{otherwise},
    \end{cases}
\end{equation*}
where $R_\mathrm{sell} \in [0,1)$ is the ratio between selling and buying price. Since we minimize over the electricity cost, we can use a losslessly relaxed convex version~\cite{BorsboomFahdzyanaEtAl2021}:
\begin{equation}\label{eq:Pg_relaxed}
    C_{\mathrm{el}}^i [k]\geq 
    \begin{cases}
    p_{\mathrm{da}}[k]\cdot \Delta t\ \cdot P_{\mathrm{g}}^i[k]\\ 
    R_\mathrm{sell} \cdot p_{\mathrm{da}}[k]\cdot \Delta t \cdot P_{\mathrm{g}}^i[k],
    \end{cases}
\end{equation}
where the grid power is constrained by:
\begin{equation}\label{eq:Pglims}
	{P}_{\mathrm{g}}^i[k] \in \left[ \underline{P}_{\mathrm{g}}^i, \overline{P}_{\mathrm{g}}^i \right].
\end{equation}

\subsection{Solar Plant}
We consider the PV power after accounting for losses $\mathbf{P}_{\mathrm{pv}}^i \in \mathbb{R}^{N_\mathrm{w}}$ as an exogenous variable to our model.

\vspace{-3pt}
\section{Centralized Optimization}\label{centralized}
In this section, we formulate the optimization problem of operating the Charging Energy Hub Network. We aim to optimize the operation of the network over all time steps in $\mathcal{K}$. We design the objective function considering the total cost of electricity and regularization terms on all $\mathbf{P}_{\mathrm{ev},j}^i$  and $\mathbf{P}_{\mathrm{dc},c}$, which aim to produce smooth power profiles.
\begin{equation*}
	J(\mathbf{C}_\mathrm{el}^i, \{\mathbf{P}_{\mathrm{ev},j}^i\}_{j\in\mathcal{S}^i}) = \mathbf{C}_{\mathrm{el}}^i \mathds{1} + \alpha_\mathrm{ev}\cdot\!\! \sum_{j\in\mathcal{S}^i} \left \lVert \mathbf{P}_{\mathrm{ev},j}^i \right \rVert^2_2 ,
\end{equation*}
where $\alpha_\mathrm{ev}$  is the regularization weight. For the hub connections we define the regularization term
\begin{equation*}
	g\left(\mathbf{P}_{\mathrm{dc},c}\right) =  \left \lVert \mathbf{P}_{\mathrm{dc},c}\right \rVert^2_2.
\end{equation*}

For comparison, we introduce two variants of the optimization problem. First, we formulate the V1G Optimization Problem that optimizes the EV charging profiles.
\begin{prob}[V1G Centralized Optimization]\label{prob:central}
	\begin{equation*}
		\begin{aligned}
			\!\min _{\substack{\{\mathbf{P}_\mathrm{ib}^i, \{\mathbf{P}_{\mathrm{ev},j}^i\}_{j\in\mathcal{S}^i}\}_i,\\ \{\mathbf{P}_{\mathrm{dc},c}\}_c}} \quad& \sum_{i\in\mathcal{H}} J(\mathbf{C}_\mathrm{el}^i, \{\mathbf{P}_{\mathrm{ev},j}^i\}_{j\in\mathcal{S}^i}) \quad + \\
			&\quad \alpha_\mathrm{dc} \cdot \sum_{c\in\mathcal{C}} g\left(\mathbf{P}_{\mathrm{dc},c}\right)\\
			\textnormal{s.t.}\quad \quad&(\ref{eq:Pbalance}-\ref{eq:Pglims})\\
			&\forall k \in \mathcal{K}, j \in \mathcal{S}^i, i \in \mathcal{H}, c\in \mathcal{C},\\
		\end{aligned}
	\end{equation*}
\end{prob}
where the optimization variables are $\mathbf{P}_{\mathrm{ev},j}^i$,  $\mathbf{P}_{\mathrm{ib}}^i \in \mathbb{R}^{N_\mathrm{w}}$ and $\mathbf{P}_{\mathrm{dc},c}\in \mathbb{R}^{N_\mathrm{w}}$, and $\alpha_\mathrm{dc}$ is a regularization weight.

Then, to obtain a baseline we do not control the EV charging power but instead constrain it to be the average session power, and optimize over the BESS and interconnection power only:

\begin{prob}[Baseline Centralized Optimization]\label{prob:central_avg_power}
	\begin{equation*}
		\begin{aligned}
			\!\min _{\substack{\{\mathbf{P}_\mathrm{ib}^i\}_i,\\ \{\mathbf{P}_{\mathrm{dc},c}\}_c}}  \quad& \sum_{i\in\mathcal{H}} J(\mathbf{C}_\mathrm{el}^i, \{\mathbf{P}_{\mathrm{ev},j}^i\}_{j\in\mathcal{S}^i}) 
			+ \alpha_\mathrm{dc} \cdot \sum_{c\in\mathcal{C}}  g\left(\mathbf{P}_{\mathrm{dc},c}\right) \\
			\textnormal{s.t.} \quad\quad&  (\ref{eq:Pbalance}-\ref{eq:Pblims}), (\ref{eq:chargingPower}-\ref{eq:Pglims})\\
			& {P}_{\mathrm{ev},j}^i [k] = \frac{E_{\mathrm{c}, j}^i}{\mathbf{a}_j^i \mathds{1} \cdot \Delta t} \;\; \forall k \in [k_{0, j}^i, k_{\mathrm{f}, j}^i],\\
			& \forall k \in \mathcal{K}, j \in \mathcal{S}^i, i \in \mathcal{H}, c\in \mathcal{C}.
		\end{aligned}
	\end{equation*}
\end{prob}

\vspace{-10pt}
\section{Distributed Optimization}\label{distributed}
\vspace{-10pt}
In this section, we describe the consensus distributed formulation of the optimal operation of the Charging Energy Hub Network. Every hub keeps their local parameters and variables private and solves a local optimization problem. Thereafter, they communicate their local versions of the shared variables $\mathbf{P}_{\mathrm{dc},c}$ to a central collector, that aims to achieve consensus by minimizing the difference between local versions of the same shared variable. To this end, we formulate a separable distributed version of Problem~\ref{prob:central} as shown in Problem~\ref{prob:distributed}.

We differentiate between the private and public views of $\mathbf{P}_{\mathrm{dc},c}$. The private version seen at hub $i$ is defined as $\mathbf{P}_{\mathrm{dc},c}^i$, whereas the public version is $\tilde{\mathbf{P}}_{\mathrm{dc},c}$. We introduce shorthand notation for the local objective function at hub $i$:
 \begin{equation*}
 	J_\mathrm{d}(\{\mathbf{P}_{\mathrm{dc},c}^i\}_{c:i\in c}) = J(\mathbf{C}_\mathrm{el}^i, \{\mathbf{P}_{\mathrm{ev},j}^i\}_{j\in\mathcal{S}^i}),
 \end{equation*}
 where $\mathbf{C}_\mathrm{el}^i$ depends on $\mathbf{P}_{\mathrm{dc},c}^i$ through~\eqref{eq:Pbalance}. The interconnection regularization term only depends on public variables:
 \begin{equation*}
 	g(\tilde{\mathbf{P}}_{\mathrm{dc},c}) = \left \lVert \tilde{\mathbf{P}}_{\mathrm{dc},c} \right \rVert_2^2.
 \end{equation*} 
We formulate the separable optimization problem
\begin{prob}[V1G Distributed Optimization]\label{prob:distributed}
	\begin{align}
			\nonumber\min _{\substack{\{\mathbf{P}_\mathrm{ib}^i, \{\mathbf{P}_{\mathrm{ev},j}^i\}_{j\in\mathcal{S}^i}\}_i, \\ \left\{\{\mathbf{P}_{\mathrm{dc},c}^i\}_i, \tilde{\mathbf{P}}_{\mathrm{dc},c}\right\}_c }} \quad& \sum_{i\in\mathcal{H}} J_\mathrm{d}\left(\{\mathbf{P}_{\mathrm{dc},c}^i\}_{c:i\in c}\right) \quad + \\
			\nonumber& \quad  \alpha_\mathrm{dc}\cdot  \sum_{c\in\mathcal{C}} g(\tilde{\mathbf{P}}_{\mathrm{dc},c}) \\
			\textnormal{s.t.} \quad\quad\quad& \mathbf{P}_{\mathrm{dc},c}^i = \tilde{\mathbf{P}}_{\mathrm{dc},c} \;\; \forall i: i\in c, \; \forall c\in\mathcal{C} \label{eq:sharedVarConstraint}\\
			\nonumber& (\ref{eq:Pbalance}-\ref{eq:Pglims}),\\
			\nonumber& \forall k \in \mathcal{K}, j \in \mathcal{S}^i,  i \in \mathcal{H}, c\in \mathcal{C},
	\end{align}
\end{prob}
which is equivalent to Problem~\ref{prob:central}.

We solve Problem~\ref{prob:distributed} using the scaled form of ADMM based on the methodology in \cite{BoydParikhEtAl2011}, for which we separate the problem into subproblems for every hub, where we augment the local objective functions with a penalty on the non-satisfaction of \eqref{eq:sharedVarConstraint}. This approach preserves hub privacy by only requiring each hub $i$ to provide information on $\mathbf{P}_{\mathrm{dc},c}^i$ and the dual value associated with \eqref{eq:sharedVarConstraint}. Then, we iteratively perform the following procedure until convergence on the shared variables:

\noindent\emph{Private variable update:} At iteration step $\ell$, we solve
\begin{equation*}
	\begin{aligned}
		\left\{\mathbf{P}_{\mathrm{dc},c}^{i,\ell+1}\right\}_{\!\! c:i\in c}
		&= \operatorname*{arg\,min}_{\{\mathbf{P}_{\mathrm{dc},c}^i\}_{c:i\in c}}
		\left(
		J_\mathrm{d}\left(\left\{\mathbf{P}_{\mathrm{dc},c}^{i}\right\}_{\!\!c:i\in c}\right) \quad  + \phantom{\sum_{c:i\in c}}\right.\\
		&\quad\left.
		\frac{\rho}{2}\!\cdot\!\sum_{c:i\in c}
			\begin{Vmatrix}	\mathbf{P}_{\mathrm{dc},c}^i - \tilde{\mathbf{P}}_{\mathrm{dc},c}^\ell
								+ \frac{\mathbf{y}^{i,\ell}_c}{\rho} 
			\end{Vmatrix}_2^2
		\right) \;\; \forall i \in \mathcal{H},
	\end{aligned}
\end{equation*}
where $\mathbf{y}^{i,\ell}_c \in \mathbb{R}^{N_\mathrm{w}}$ is the dual variable at iteration step $\ell$ and $\rho$ is the augmented Lagrangian coefficient.

\noindent\emph{Public variable update:} Next, in the collector step, we retrieve the solution $\mathbf{P}_{\mathrm{dc},c}^{i, \ell+1}$ from each local problem and update the public variable as
\begin{equation*}
	\tilde{\mathbf{P}}_{\mathrm{dc},c}^{\ell+1} = \frac{\rho}{2\!\cdot\! (\rho+ \alpha_\mathrm{dc})}\!\cdot\! \sum_{i:i\in c} \left( \mathbf{P}_{\mathrm{dc},c}^{i,\ell+1} + \frac{\mathbf{y}^{i,\ell}_c}{\rho}\right) \quad \forall c \in \mathcal{C},
\end{equation*}
where we use $i:i\in c$ to denote each hub $i$ connected by edge $c$. To respect the feasibility of each element of $\tilde{\mathbf{P}}_{\mathrm{dc},c}^{\ell+1}$ at every iteration step, we include a saturation function:
\begin{equation*}
	\tilde{P}_{\mathrm{dc},c}^{\ell+1} [k] = \mathrm{sat}_{\left[\underline{P}_\mathrm{dc},\overline{P}_\mathrm{dc}\right]} \tilde{P}_{\mathrm{dc},c}^{\ell+1} [k]\;\; \forall k\in\mathcal{K},\;\forall c\in\mathcal{C}.
\end{equation*}

\noindent\emph{Dual variable ascent:} In the last step, we perform $\forall i \in \mathcal{H}$:
\begin{equation*}
	\mathbf{y}^{i,\ell+1}_c = \mathbf{y}^{i,\ell}_c + \rho \cdot \left( \mathbf{P}_{\mathrm{dc},c}^{i, \ell+1} - \tilde{\mathbf{P}}_{\mathrm{dc},c}^{\ell+1} \right)  \quad \forall c:i\in c.
\end{equation*}

\noindent\emph{Stopping criteria:} Following the variable updates at each iteration step $\ell$, we check the stopping criteria defining the acceptable consensus among hubs. We implement the stopping criteria following the methodology in~\cite{BoydParikhEtAl2011}. We use the sum of squares of the primal residual $\mathbf{r}$
\begin{equation*}
	\left \lVert \mathbf{r}^\ell \right \rVert_2^2 = \sum_{i\in\mathcal{H}} \sum_{c:i\in c} \left \lVert  \mathbf{P}_{\mathrm{dc},c}^{i,\ell} - 	\tilde{\mathbf{P}}_{\mathrm{dc},c}^\ell \right \rVert_2^2,
\end{equation*}
and the sum of squares of the dual residual $\mathbf{s}$
\begin{equation*}
	\left \lVert  \mathbf{s}^\ell \right \rVert_2^2 = N_\mathrm{h}\cdot\rho^2\cdot\sum_{c:i \in c}\left \lVert \tilde{\mathbf{P}}_{\mathrm{dc},c}^{\ell+1} - 	\tilde{\mathbf{P}}_{\mathrm{dc},c}^\ell \right \rVert_2^2.
\end{equation*}
The primal stopping criteria  takes place when $\left \lVert  \mathbf{r}^\ell \right \rVert_2 <~\varepsilon_\mathrm{p}^\ell$, in which the threshold is defined as
\begin{equation*}
	\varepsilon_\mathrm{p}^\ell =  \sqrt{p}\cdot\varepsilon_\mathrm{abs} + \varepsilon_\mathrm{rel}\cdot \max\left\{ \left \lVert  \left[\mathbf{P}\right]_\mathrm{dc}^\ell \right \rVert_\mathrm{F}, \left \lVert \left[\tilde{\mathbf{P}}\right]_\mathrm{dc, c}^\ell \right \rVert_\mathrm{F} \right\},
\end{equation*}
where we use $\left[\mathbf{P}\right]$ to denote a vector of all stacked time series of a certain variable, e.g., $\left[\mathbf{P}\right]^{\ell}_\mathrm{dc} = \{\mathbf{P}_{\mathrm{dc},c}^{i,\ell}\}_{i\in \mathcal{H},c \in \mathcal{C}}$, $p$ is the number of elements in $\left[\mathbf{P}\right]_\mathrm{dc} $, $\varepsilon_\mathrm{abs}$ and $\varepsilon_\mathrm{rel}$ are fixed tolerances and $\left \lVert  .\right \rVert_\mathrm{F}$ refers to the Frobenius norm. Similarly, we reach the stopping criteria of the dual  when $\left \lVert  \mathbf{s}^\ell \right \rVert_2 <~\varepsilon_\mathrm{d}^\ell$, in which the threshold is defined as
\begin{equation*}
	\varepsilon_\mathrm{d}^\ell = \sqrt{n}\cdot\varepsilon_\mathrm{abs} + \varepsilon_\mathrm{rel} \cdot \left \lVert  \left[\mathbf{y}\right]^\ell \right \rVert_\mathrm{F},
\end{equation*}
where $n$ is the number of elements in the stacked dual variable $\left[\mathbf{y}\right]$.

%% file: Sections/results.tex
\section{Case Study Results}
\vspace{-4pt}

\begin{table}[t]
	\vspace{10pt}
	\centering
	\caption{Charging Energy Hub Network specification. 
}
	\label{tab:charging_energy_hub_specs}
	\begin{tabularx}{\columnwidth}{lXl}
		\toprule
		\textbf{Parameters} & \textbf{Values} \\
		\midrule
		\textbf{BESS}\\
		$\underline{E}_{\mathrm{b}}^{1}, \underline{E}_{\mathrm{b}}^{2}$ & $100~[\mathrm{kWh}]$ \\
		$\overline{E}_{\mathrm{b}}^{1}\,\overline{E}_{\mathrm{b}}^{2}$ &$900~[\mathrm{kWh}]$ \\
		$\underline{P}_{\mathrm{b}}^{1},\underline{P}_{\mathrm{b}}^{2}$ & $-300~[\mathrm{kW}]$ \\
		$\overline{P}_{\mathrm{b}}^{1},\overline{P}_{\mathrm{b}}^{2}$ & $300~[\mathrm{kW}]$ \\
		$\eta^{1},\,\eta^{2}$ & $0.95$ \\
		\textbf{PV Rated Power}\\
		$\overline{P}_{\mathrm{pv}}^{1},\,\overline{P}_{\mathrm{pv}}^{2}$ & $1,\,0.5~[\mathrm{MW}]$ \\
		\textbf{EV}\\
		$\overline{N}_{\mathrm{ev}}^{1},\,\overline{N}_{\mathrm{ev}}^{2},\,\overline{N}_{\mathrm{ev}}^{3}$ & $130,\,40,\,40$ \\
		$\underline{P}_{\mathrm{ev}}$ & $0~[\mathrm{kW}]$ \\
		$\overline{P}_{\mathrm{ev}}$ & $300~[\mathrm{kW}]$ \\
		\textbf{Distribution Grid}\\
		$\underline{P}_{g}^{1},\,\underline{P}_{g}^{3}$ & $-2,\,-1~[\mathrm{MW}]$ \\
		$\overline{P}_{g}^{1},\,\overline{P}_{g}^{3}$ & $2,\,1~[\mathrm{MW}]$ \\
		$R_\mathrm{sell}$ & 0.9 \\
		\textbf{Interconnection}\\
		$\underline{P}_{\mathrm{dc}}$ & $-1.2~[\mathrm{MW}]$ \\
		$\overline{P}_{\mathrm{dc}}$ & $1.2~[\mathrm{MW}]$ \\
		\bottomrule
	\end{tabularx}
\end{table}

For this case study, we use the three hub network presented in Fig.~\ref{fig:network} ($\mathcal{H}=\{1, 2, 3\},\, \mathcal{C}=\{(1,2),(1,3)\}$) and include all component details in Tab.~\ref{tab:charging_energy_hub_specs}. Hub 1 contains all CEH components, while hub 2 lacks a main grid connection and hub 3, a BESS component and PV panels. We use a real life dataset of EV users in the Netherlands, provided by Maxem, a company working on power and energy management of EV charging stations. In addition, for hub 2 and 3, we reduce the original EV charging demand by downsampling the number of sessions per evaluation day to $\overline{N}_\mathrm{ev}^2$ and $\overline{N}_\mathrm{ev}^3$. We use day-ahead electricity prices and emission factor of the electricity in the Netherlands from the European Network of Transmission System Operators for Electricity (ENTSO-E) \cite{entsoeAPI2024}. For the PV production, we use historical weather information in the Netherlands from the Open-Meteo.com Weather API \cite{openmeteoAPI2024} and obtain the solar power following the model in \cite{IzadiFernandezZapicoEtAl2026}. The evaluation set includes a total of 30 days in 2025, composed of three days per month available. We use a step resolution $\Delta t = \unit[15]{min}$, a two-day ahead optimization horizon ($N_\mathrm{w}=192$) and initialize each evaluation day at 11:00. During evaluation, we assume perfect knowledge of $\mathbf{P}_\mathrm{pv}^i$, $\mathbf{p}_\mathrm{da}$, $E_\mathrm{c}^i$, $\mathbf{a}^i_j$. In previous work, we investigated the influence of predicted inputs in the online operation of the energy hub \cite{FernandezZapicoHofmanEtAl2025}.  We set all regularization terms to $10^{-3}$. We provide more details on the public repository of this work \footnote{\url{{https://github.com/diegofz/CEHNetworks_ADMM}}}.
\vspace{-5pt}
\begin{figure}[h]
	\centering
	\includegraphics[width=0.40\textwidth]{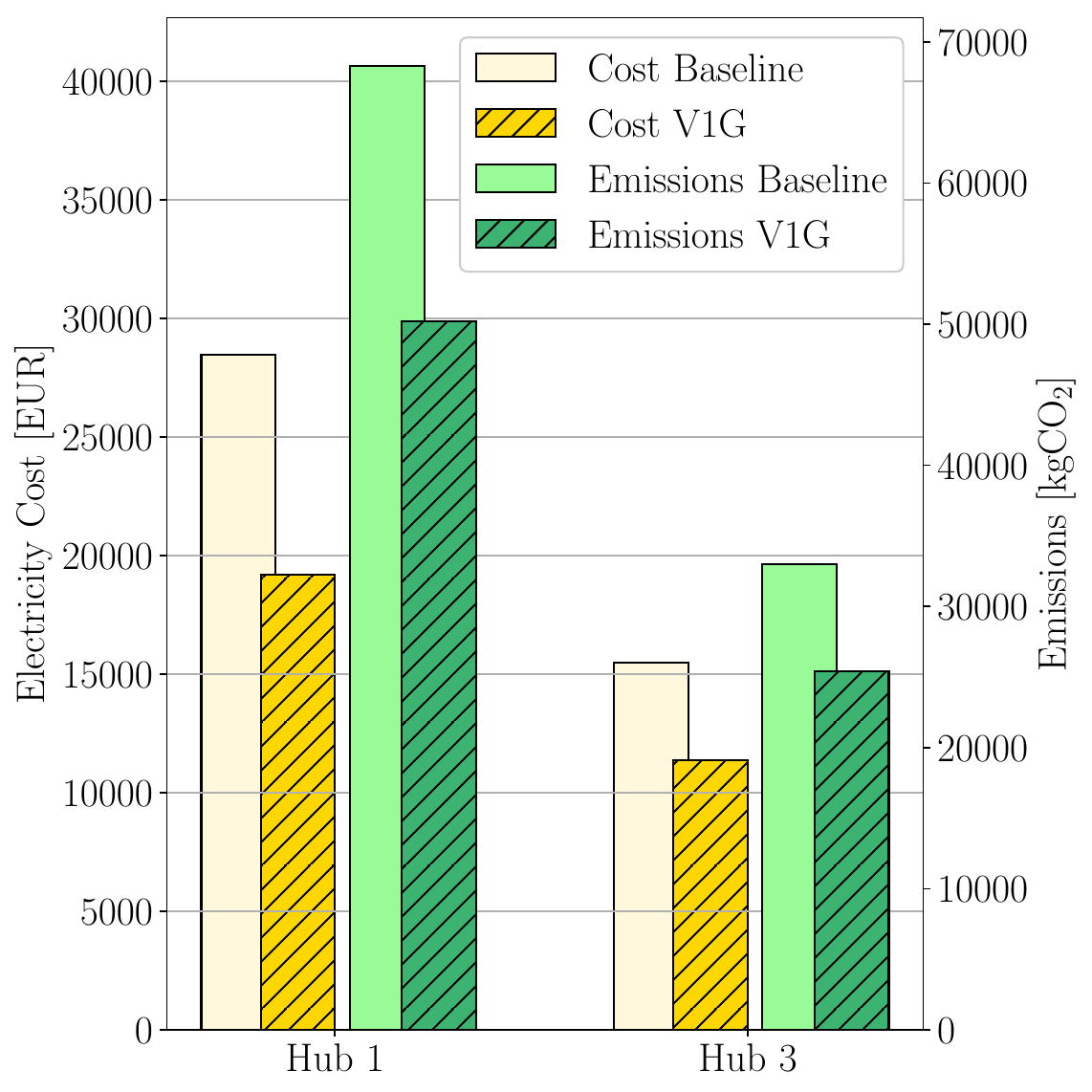}
	\caption{Total Electricity Cost and Emissions comparison between Problem~\ref{prob:central} (V1G) and~{\ref{prob:central_avg_power}} (Baseline) in a 30-day evaluation.}
	\label{fig:comparison}
\end{figure}

\subsection{V1G Performance}
\begin{figure}[h]
	\centering
	\includegraphics[width=0.95\columnwidth]{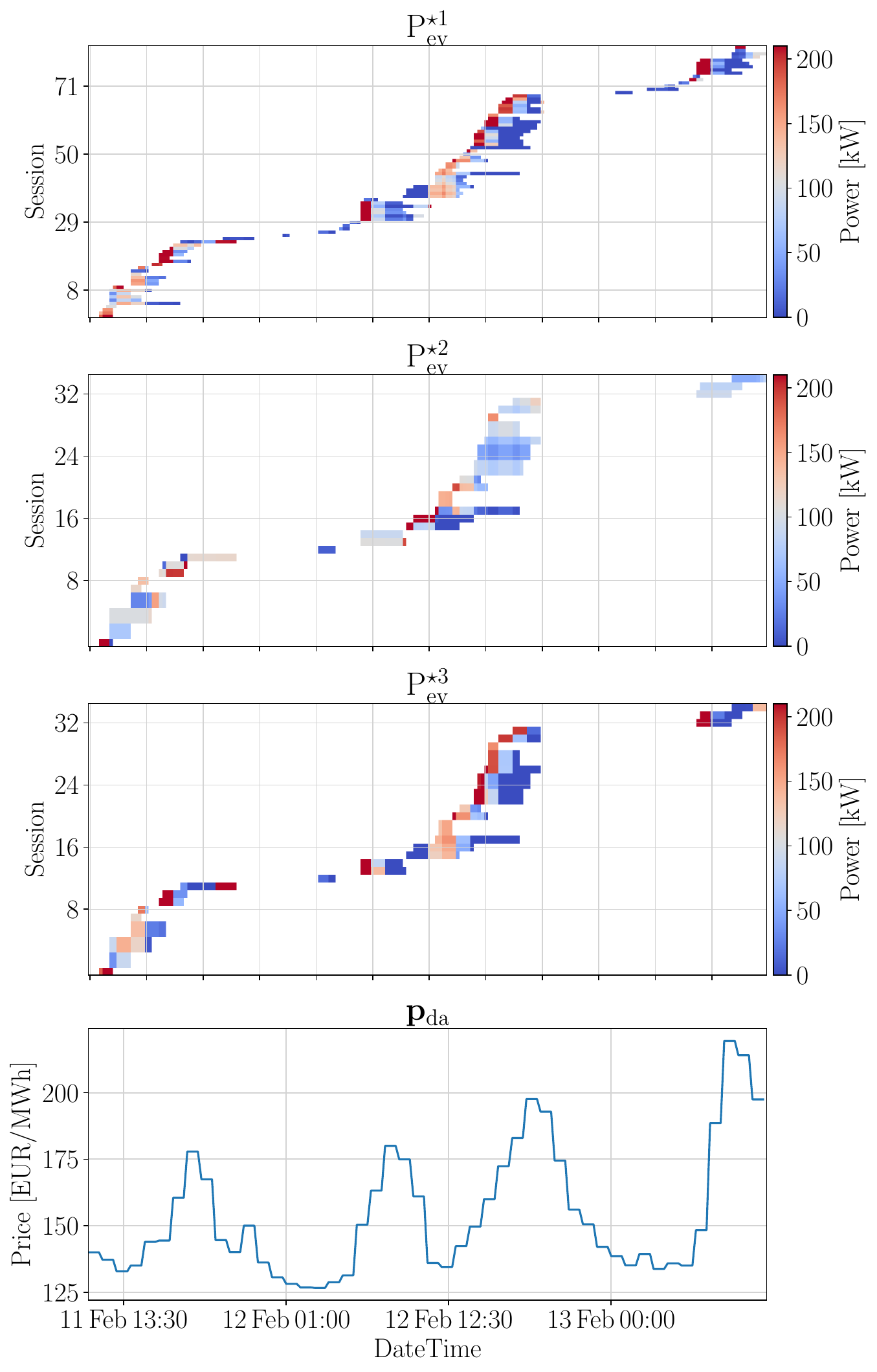}
	\caption{Optimal V1G EV charging profiles for hub $1$ ($P_\mathrm{ev}^{\star 1}$), $2$ ($P_\mathrm{ev}^{\star 2}$) and $3$ ($P_\mathrm{ev}^{\star 3}$), and $\mathbf{p}_\mathrm{da}$.}
	\label{fig:pev}
\end{figure}

This section compares the control performance of the V1G Optimization in Problem~\ref{prob:central} with the Baseline Optimization Problem~\ref{prob:central_avg_power}. In Tab.~\ref{tab:cost}, we report  a 30\% reduction in the total electricity cost and a 25\% reduction in the total electricity emissions during evaluation, when comparing V1G to the Baseline. In Fig.~\ref{fig:comparison}, we do a per hub breakdown of the total costs and find a greater cost and emission reduction in hub 1, which has a higher capacity to sell power to the main grid. We show the optimal EV charging profiles for each hub ($P_\mathrm{ev}^{\star i}$) during a sample day in Fig.~\ref{fig:pev}. Furthermore, we include $\mathbf{p}_\mathrm{da}$ which motivates higher EV charging power when there are lower within-session prices.

\begin{table}[h]
	\vspace{10pt}
	\centering
	\caption{Absolute and Normalized Total Electricity Cost ($C_\mathrm{el}$) and Emissions ($E_\mathrm{el}$) of V1G and Baseline in a 30-day evaluation.}
	\label{tab:cost}
	\begin{tabular}{c|c|c|c|c}
		Version & $C_{\mathrm{el}}$[$10^3$ EUR] & $C_{\mathrm{el}}$[\%] & $E_{\mathrm{el}}$[$10^3$ kgCO$_2$] & $E_{\mathrm{el}}$[\%]\\ \hline
		Baseline & 43.9 & 100 & 101.3&100\\
		V1G & 30.5 & 70 &75.6 & 75\\ \hline
	\end{tabular}
\end{table}


\vspace{-4pt}
\subsection{ADMM Convergence}
\begin{figure}[H]
	\centering
	\includegraphics[width=0.8\columnwidth]{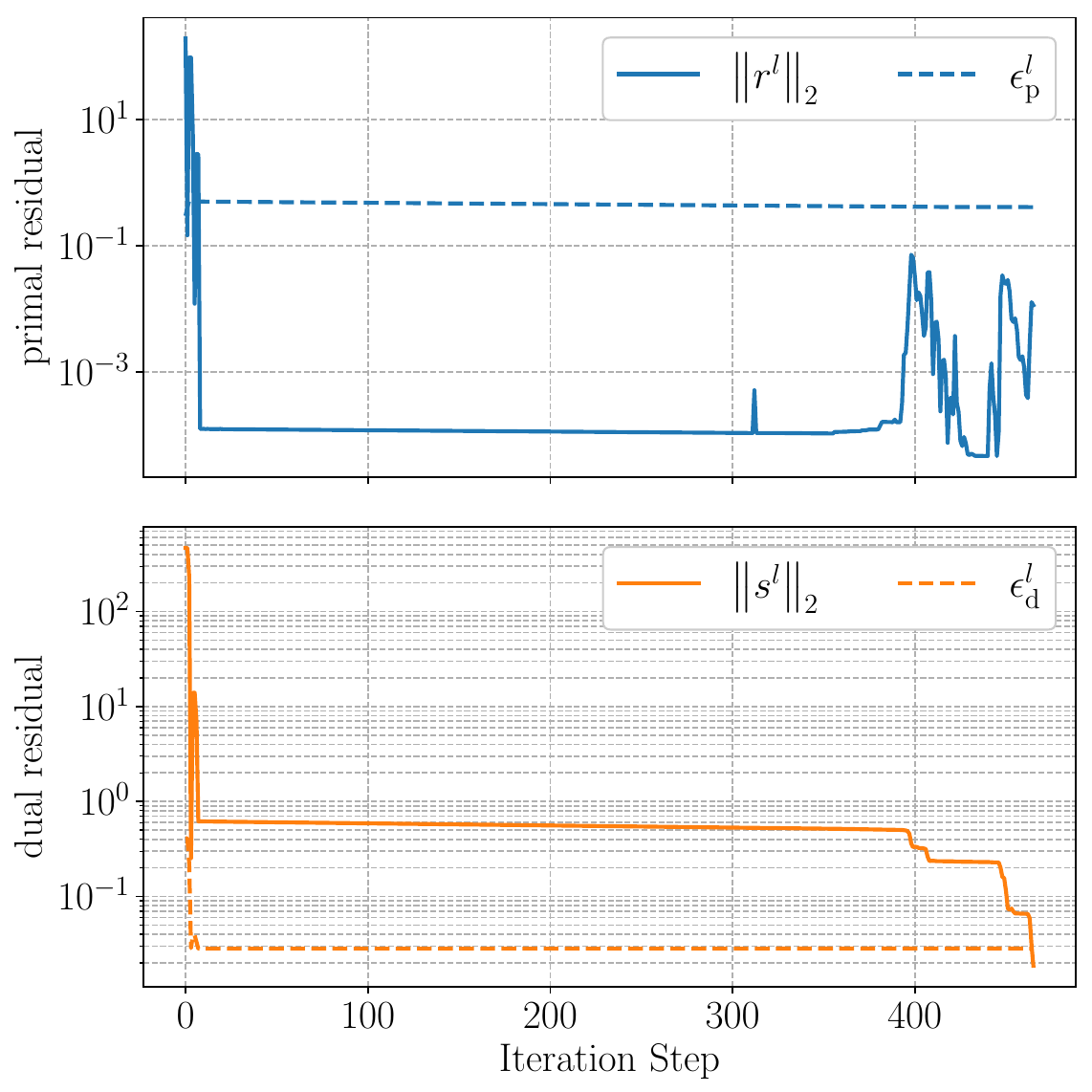}
	\caption{
		ADMM convergence at the stopping criteria of the primal ($\left \lVert  r^l \right \rVert_2~<\varepsilon_\mathrm{p}^l$) and dual ($\left \lVert  s^l \right \rVert_2 ~< \varepsilon_\mathrm{d}^l$) residuals for the Distributed V1G in an evaluation day.
	}
	\label{fig:convergence}
\end{figure}

\begin{figure*}[!t]
	\centering
	\includegraphics[width=0.80\textwidth]{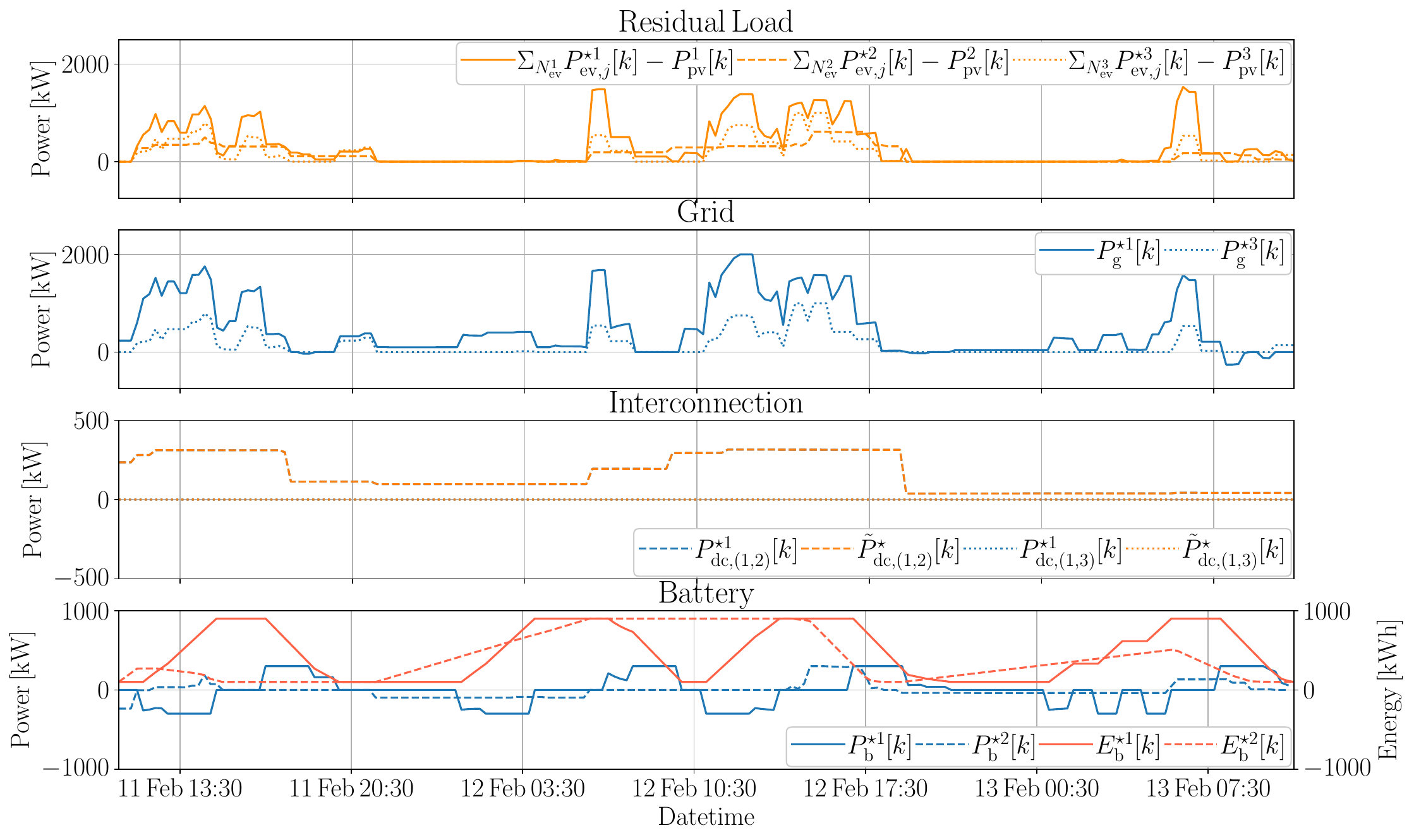}
	 \vspace{-2pt}
	\captionof{figure}{Optimal two-day ahead operation of the Charging Energy Hub Network using ADMM to solve the V1G Distributed Optimization.  For each hub $i$, at each time step $k$, we present the optimal residual load ($\sum_{j=1}^{N_\mathrm{ev}^i}P_{\mathrm{ev}, j}^{\star i}[k] - P_{\mathrm{pv}}^i[k]$). We also include the local optimal variables ($P_{\mathrm{g}}^{\star i}[k]$, $P_{\mathrm{b}}^{\star i}[k]$, $E_{\mathrm{b}}^{\star i}[k]$). For the interconnections, we plot the optimal public variable $\tilde{P}_{\mathrm{dc},c}^\star[k]$ for each edge $c$ and the optimal private variables from hub $1$ $P_{\mathrm{dc},(1,2)}^{\star 1}[k]$, $P_{\mathrm{dc},(1,3)}^{\star 1}[k]$. }
	\label{fig:main_res}
\end{figure*}

In this section we present the results of solving the distributed optimization formulated in Problem~\ref{prob:distributed} using ADMM. Setting $\rho = 2$ and $\varepsilon_\mathrm{rel} = \varepsilon_\mathrm{abs} = 10^{-3}$, we report the stopping criteria of the ADMM method for an evaluation day in Fig.~\ref{fig:convergence}. We show the optimal operation of the CEH Network for the evaluation day in Fig.~\ref{fig:main_res}, including all component variables involved. Finally, we report the accuracy of the ADMM solution in Tab.~\ref{tab:convergence} based on the normalized mean absolute error on evaluation day $d$:
\begin{equation*}
	\mathrm{nMAE}^d = \frac{1}{N_Y\cdot \overline{Y}}\cdot \sum_{w=1}^{N_Y} |Y_w^d-\hat{Y}_w^d|,
\end{equation*}
where $Y_w^d$ is the value of element $w$ in variable $Y$ for day $d$ of the solution of Problem~\ref{prob:central} and, similarly, $\hat{Y}_w^d$ is the corresponding value of the solution of Problem~\ref{prob:distributed}. The element $ w \in \{ 1, ..., N_Y\}$, where $N_Y$ is the number of elements in variable $Y$. We normalize using $\overline{Y}$, which is the corresponding maximum value presented in Tab.~\ref{tab:charging_energy_hub_specs}. For each optimization variable, we consider the $\mathrm{nMAE}^d$ and report the average ($\mathrm{nMAE_{avg}}$), maximum ($\mathrm{nMAE_{max}}$) and minimum ($\mathrm{nMAE_{min}}$) value. This shows that, on average, at day $d$, the ADMM method achieves results with a $\mathrm{nMAE}^d<~1\%$ for all optimization variables. In the worst case, an error of $2.4\%$ takes place for the BESS power injections on hub $2$, during this evaluation.

\begin{table}[H]
	\vspace{10pt}
	\centering
	\caption{Average, maximum and minimum daily Normalized Mean Absolute Error [\%] between Centralized  and Distributed V1G in a 30-day evaluation.}
	\label{tab:convergence}
	\begin{tabular}{c|c|c|c}
		Variables & $\mathrm{nMAE_{avg}}$ & $\mathrm{nMAE_{max}}$ & $\mathrm{nMAE_{min}}$ \\ \hline
		$\mathbf{P}_\mathrm{ev}^{\star 1}, \mathbf{P}_\mathrm{ev}^{\star 2}, \mathbf{P}_\mathrm{ev}^{\star 3}$  & 0.0, 0.0, 0.0&	0.04& 0\\
		$\mathbf{P}_\mathrm{g}^{\star 1}, \mathbf{P}_\mathrm{g}^{\star 3}$ &	0.16, 0.01&	0.4&	0.0\\
		$\tilde{\mathbf{P}}_{\mathrm{dc},(1,2)}^\star, \tilde{\mathbf{P}}_{\mathrm{dc},(1,3)}^\star$ & 0.26, 0.01 & 0.6 &	0\\
		$\mathbf{P}_\mathrm{b}^{\star 1}, \mathbf{P}_\mathrm{b}^{\star 2}$  &  0.21, 0.94 &	2.4&	0.05\\
		\hline
	\end{tabular}
\end{table}

%% file: Sections/conclusions.tex
\section{Conclusion}
\vspace{-4pt}
In this work, we presented optimization models to study the operation of a network of charging energy hubs.
Considering vehicle charging profiles as optimization variables, we showed that smart scheduling can substantially reduce energy costs compared to fixed-power profiles.
Moreover, our convex problem formulation not only guaranteed global optimality of the solutions, but enabled an effective problem decomposition that allows to solve it for large networks in a distributed fashion via standard ADMM algorithms that preserve privacy.

In future work, we would like to implement the current approach in a receding horizon fashion with stochastic predictions. Moreover, we would like to leverage the privacy preserving features of the proposed distributed optimization algorithm to study coordination of multiple hub owners.

%

%% file: SML_Papers.bib
@Preamble{"\newcommand{\noopsort}[1]{} " #
"\newcommand{\printfirst}[2]{#1} " #
"\newcommand{\singleletter}[1]{#1} " #
"\newcommand{\switchargs}[2]{#2#1} "}

@String{jrn_Elsevier_AE                = {{Applied Energy}}}

@String{jrn_IEEE_CSL                   = {{IEEE Control Systems Letters}}}

@String { jrn_IEEE_TII              = {{IEEE Transactions on Industrial Informatics}} }

@String { jrn_IEEE_TITS             = {{IEEE Transactions on Intelligent Transportation Systems}} }

@String { jrn_IEEE_TSG              = {{IEEE Transactions on Smart Grid}} }

@String{jrn_IEEE_TTE                   = {{IEEE Transactions on Transportation Electrification}}}

@String{jrn_IEEE_TVT                   = {{IEEE Transactions on Vehicular Technology }}}

@String { proc_IEEE_CDC             = {{Proc.\ IEEE Conf.\ on Decision and Control}} }

@String{proc_IEEE_ESTS                 = {{IEEE Electric Ship Technologies Symposium}}}

@String { proc_IFAC_WC              = {{IFAC World Congress}} }

@Article{VehlhaberSalazar2023b,
  author    = {Vehlhaber, F. and Salazar, M.},
  title     = {Electric Aircraft Assignment, Routing, and Charge Scheduling Considering the Availability of Renewable Energy},
  journal   = jrn_IEEE_CSL,
  year      = {2023},
  volume    = {7},
  pages     = {3669-3674},
  note      = {Available online at \url{http://arxiv.org/pdf/2309.09793v1}},
  doi       = {10.1109/lcsys.2023.3339998},
}

@Article{BorsboomFahdzyanaEtAl2021,
  author    = {Borsboom, O. and Fahdzyana, C. A. and Hofman, T. and Salazar, M.},
  title     = {A Convex Optimization Framework for Minimum Lap Time Design and Control of Electric Race Cars},
  journal   = jrn_IEEE_TVT,
  volume    = {70},
  number    = {9},
  pages     = {8478--8489},
  year      = {2021},
  doi       = {\,10.1109/TVT.2021.3093164},
}

@InProceedings{IzadiFernandezZapicoEtAl2026,
author    = {Izadi, M. and Fernandez Zapico, D. and Salazar, M. and Hofman, T.},
title     = {Optimal Sizing of Charging Energy Hubs for Heavy-Duty Electric Transport through Co-Optimization},
booktitle = proc_ifac_WC,
year      = {2026},
note      = {In Press},
}

@InProceedings{BertucciHofmanEtAl2025,
author    = {Bertucci, J. P. and Hofman, T. and Salazar, M.},
title     = {Simultaneous Optimization of Electric Ferry Operations and Charging Infrastructure},
booktitle = proc_IEEE_ESTS,
year      = {2025},
}

@InProceedings{FernandezZapicoHofmanEtAl2025,
  author    = {Fernandez Zapico, D. and Hofman, T. and Salazar, M.},
  title     = {Stochastic Model Predictive Control of Charging Energy Hubs with Conformal Prediction},
  booktitle = proc_ieee_cdc,
  year      = {2025},
}


%% file: energyhubs.bib
@String{jrn_Elsevier_AE = {{Applied Energy}}}

@String{jrn_FTML = {{Found. Trends Mach. Learn.}}}

@String{jrn_IEEE_TII      = {{IEEE Trans.\ on Industrial Informatics}}}

@String{jrn_IEEE_TITS     = {{IEEE Transactions on Intelligent Transportation Systems}}}

@String { jrn_IEEE_TSG  = {{IEEE Transactions on Smart Grid}} }

@String{jrn_IEEE_TTE      = {{IEEE Transactions on Transportation Electrification}}}

@String{jrn_IEEE_TVT                   = {{IEEE Trans.\ on Vehicular Tech.\ }}}

@String{proc_IEEE_CDC_52  = {{Proc.\ IEEE Conf.\ on Decision and Control}}}

@Online{entsoeAPI2024,
	author       = {{European Network of Transmission System Operators for Electricity}},
	organization = {{ENTSO-E}},
	title        = {{Transparency Platform}},
	year         = {2024},
	url          = {https://transparency.entsoe.eu/},
}

@Online{openmeteoAPI2024,
	author       = {Zippenfenig, P.},
	title        = {{Open-Meteo.com Weather API}},
	year         = {2023},
	url          = {https://open-meteo.com},
}

@Article{MotlaghOladigboluEtAl2025,
  author    = {Motlagh, S. G. and Oladigbolu, J. and Li, L.},
  title     = {A review on electric vehicle charging station operation considering market dynamics and grid interaction},
  journal   = jrn_Elsevier_AE,
  volume    = {392},
  pages     = {126058},
  year      = {2025},
}

@Article{NimalsiriMediwaththeEtAl2019,
  author    = {Nimalsiri, N. I. and Mediwaththe, C. P. and Ratnam, E. L. and Shaw, M. and Smith, D. B. and Halgamuge, S. K.},
  title     = {A survey of algorithms for distributed charging control of electric vehicles in smart grid},
  journal   = jrn_IEEE_TITS,
  volume    = {21},
  number    = {11},
  pages     = {4497--4515},
  year      = {2019},
}

@Article{LeFlochBansalEtAl2019,
  author    = {Le Floch, C. and Bansal, S. and Tomlin, C. J. and Moura, S. J. and Zeilinger, M. N.},
  title     = {Plug-and-play model predictive control for load shaping and voltage control in smart grids},
  journal   = jrn_IEEE_TSG,
  volume    = {10},
  number    = {3},
  pages     = {2334--2344},
  year      = {2019},
}

@Inproceedings{RiveraWolfrumEtAl2013,
  author    = {Rivera, J. and Wolfrum, P. and Hirche, S. and Goebel, C. and Jacobsen, H.-A.},
  title     = {Alternating direction method of multipliers for decentralized electric vehicle charging control},
  booktitle = proc_IEEE_CDC_52,
  year      = {2013},
  pages     = {6960--6965},
}

@Article{KianiSheshyekaniEtAl2024,
  author    = {Kiani, S. and Sheshyekani, K. and Dagdougui, H.},
  title     = {{ADMM}-based hierarchical single-loop framework for {EV} charging scheduling considering power flow constraints},
  journal   = jrn_IEEE_TTE,
  volume    = {10},
  number    = {1},
  pages     = {1089--1100},
  year      = {2024},
}

@Article{MansouriNematbakhshEtAl2024,
  author    = {Mansouri, S. A. and Nematbakhsh, E. and Ramos, A. and Tostado-V{\'e}liz, M. and Aguado, J. A. and Aghaei, J.},
  title     = {A robust {ADMM}-enabled optimization framework for decentralized coordination of microgrids},
  journal   = jrn_IEEE_TII,
  volume    = {21},
  number    = {2},
  pages     = {1479--1488},
  year      = {2024},
}

@Article{BoydParikhEtAl2011,
	author = {Boyd, S. and Parikh, N. and Chu, E. and Peleato, B. and Eckstein, J.},
	title = {{Distributed Optimization and Statistical Learning via the Alternating Direction Method of Multipliers}},
	year = {2011},
	volume = {3},
	number = {1},
	journal = jrn_FTML,
	pages = {1–122},
}

@Article{YanChen2023,
	author={Yan, D. and Chen, Y.},
	journal=jrn_IEEE_TSG, 
	title={{A Distributed Online Algorithm for Promoting Energy Sharing Between EV Charging Stations}}, 
	year={2023},
	volume={14},
	number={2},
	pages={1158-1172},
	}

@Article{ZhouZouEtAl2021,
	author    = {Zhou, X. and Zou, S. and Wang, P. and Ma, Z.},
	title     = {{ADMM}-Based Coordination of Electric Vehicles in Constrained Distribution Networks Considering Fast Charging and Degradation},
	journal   = jrn_IEEE_TITS,
	volume    = {22},
	number    = {1},
	pages     = {565--578},
	year      = {2021},
}

@Article{NasiriZeynaliEtAl2024,
	author    = {Nasiri, N. and Zeynali, S. and Ravadanegh, S. N. and Kubler, S.},
	title     = {Moment-based distributionally robust peer-to-peer transactive energy trading framework between networked microgrids, smart parking lots and electricity distribution network},
	journal   = jrn_IEEE_TSG,
	volume    = {15},
	number    = {2},
	pages     = {1965--1977},
	year      = {2024},
}
